\documentclass[12pt,english]{article}
\usepackage[LGR,T1]{fontenc}
\usepackage[latin1]{inputenc}
\usepackage{geometry}
\usepackage{textcomp}
\usepackage{tipa}
\usepackage{amsmath}
\usepackage{amssymb}
\usepackage{graphicx}

\makeatletter

\DeclareRobustCommand{\greektext}{%
  \fontencoding{LGR}\selectfont\def\encodingdefault{LGR}}
\DeclareRobustCommand{\textgreek}[1]{\leavevmode{\greektext #1}}
\DeclareFontEncoding{LGR}{}{}
\DeclareTextSymbol{\~}{LGR}{126}

\@ifundefined{date}{}{\date{}}

\usepackage{units}\usepackage{amstext}\usepackage{dsfont}\usepackage{setspace}\usepackage{mathrsfs}\usepackage{longtable}\usepackage{times}\usepackage{ulem}\usepackage{babel}

\newcommand{\suppress}[1]{}

\newlength\wvtextpercent
\newbox\strikebox
\def\strike#1{\setbox\strikebox \hbox{<#1>}\hbox{\raise0.5ex\hbox to 0pt{\vrule height 0.4pt width \wd\strikebox\hss}\copy\strikebox}}

\makeatother

\usepackage{babel}
\begin{document}


\linespread{1.5}

\title{\textbf{\Huge{}On the Einstein-Podolsky-Rosen paradox using discrete
time physics}}

\author{\textbf{Roland Riek} \\
 \textbf{Laboratory of Physical Chemistry, ETH Zurich, Switzerland}}

\maketitle
Keywords:

\vspace{3cm}

\linespread{1}

Address for correspondence:\\
 Roland Riek\\
 Laboratory of Physical Chemistry\\
 ETH Zuerich\\
 Wolfgang-Pauli-Strasse 10\\
 HCI F 225\\
 CH-8093 Zurich\\
 Tel.: +41-44-632 61 39\\
 e-mail: roland.riek@phys.chem.ethz.ch

\textbf{\large{}\newpage{} }{\large \par}

\section{Abstract}

The Einstein-Podolski-Rosen paradox highlights several strange properties
of quantum mechanics including the super position of states, the non
locality and its limitation to determine an experiment only statistically.
Here, this well known paradox is revisited theoretically for a pair
of spin $\frac{1}{2}$ systems in a singlet state under the assumption
that in classical physics time evolves in discrete time steps $\Delta t$
while in quantum mechanics the individual spin system(s) evolve(s)
between the eigenstates harmonically with a period of $4\Delta t$.
It is further assumed that time is a single variable, that the quantum
mechanics time evolution and the classical physics discrete time evolution
are coherent to each other, and that the precision of the start of
the experiment and of the measurement time point are much less than
$\Delta t$. Under these conditions, it is demonstrated for a spin
$\frac{1}{2}$ system that the fast oscillation between the eigen
states spin up $|\uparrow>$ and spin down $|\downarrow>$ reproduces
the expected outcome of a single measurement as well as ensemble measurements
without the need of postulating a simultaneous superposition of the
spin system in its quantum state. When this concept is applied to
a spin $\frac{1}{2}$ system pair in a singlet state it is shown that
no entanglement between the two spins is necessary to describe the
system resolving the Einstein-Podolski-Rosen paradox.

\section{Significance Statement}

The Einstein-Podolsky-Rosen paradox is one of the most intriguing
concepts in physics that brought to attention the very strange phenomena
of entanglement in quantum mechanics, which finds experimental support
through the Bell inequalities and correspondingly set up experiments.
However, by introducing time as a single variable, which is continuous
in quantum mechanics but discrete in classical physics, time becomes
a non local hidden variable that resolves both the Bell inequalities
and the Einstein-Podolsky Rosen paradox as demonstrated here for a
two spin ½ system in a singlet state because no quantum entanglement
is required to describe the singlet state.

\section{Introduction}

The Einstein-Podolsky-Rosen paradox (EPR) (1) is a stimulating Gedankenexperiment
(thought experiment) in quantum mechanics challenging quantum mechanics
as a complete description of physical reality. Motivated by the EPR
paradox Schrödinger introduced the term entanglement (in german: Verschränkung)
(2). An entangled system is thereby defined as a system whose quantum
state cannot be described by a product of states of its individual
constituents. With other words, if a system is entangled, the state
of one constituent is not independent from the other and thus can
not be fully described without considering its counterpart. A prototypical
example thereof is the singlet state of two particles with spin $\frac{1}{2}\text{\textcrh}$
(normalized in the following) having together spin zero (3). Because
the total spin is zero, whenever the first particle is measured to
have spin up on some axis $i$ $<\uparrow_{1}|S_{i}|\uparrow_{1}>\equiv<\uparrow_{1}|\uparrow_{1}>_{i}$,
the other (when measured on the same axis) is always observed to have
spin down $<\downarrow_{2}|\downarrow_{2}>_{i}$ even upon long distance
separation. Vice versa, if the first particle is measured to have
spin down, the other will have/has spin up, respectively. It is the
paradox that before the measurement particle 1 is superimposed and
thus no decision has been made yet whether at the measurement it will
have spin up or down, but ones the measurement is performed the state
of the entire entangled system collapses instantaneously so that particle
2 has the orthogonal spin state of particle 1 without requiring any
information transfer (which requires time) between the particles.
Because such a behavior is regarded to violate causality, Einstein
and others concluded that the established formulation of quantum mechanics
must be incomplete (1). A possible solution to this dilemma appeared
to be the introduction of hidden variables, which, while not accessible
to the observer, determine the future outcome of the spin measurements
before the separation of the two particles. With other words, each
particle carries the necessary information with it upon separation,
and thus no transmission of information transfer from one particle
to the other is required at the time of measurement. However, Bell's
inequality theorem {[}4{]} rules out local hidden variables as an
explanation of the mentioned paradox and since the Bell inequalities
have been supported by experiments (5-7), EPR is considered no longer
to be a paradox and the non local nature of quantum mechanics is thus
widely accepted. It is however important to note that Bell's theorem
still permits the existence of non-local hidden variables such as
the Bohm interpretation of quantum mechanics (8), which states that
all particles in the universe are able to exchange information instantaneously,
or less demanding that there is a non-local connection of some sort
between the system under study and the measurement devices as well
as between the measurement apparatus themselves. 

In the presented approach the latter concept is realized by the assumption,
that time (unlike space coordinates) is a single variable and that
time is continuous in quantum mechanics but in classical physics evolves
in discrete steps in a coherent fashion between the two frames. This
rational guarantees that both measurement devises as well as the system
under study are in clock and thus non-locally connected. Furthermore,
it is assumed that in the quantum mechanical description of the system
there is no super-position of quantum states, but rather a fast oscillation
with a periodicity proportional to the classical time step size. This
modification to quantum mechanics allows not only to describe a quantum
system without super position, another paradox illustrated for example
by Schrödinger's cat (3, 9), but is able to resolve the EPR paradox
because the system behaves deterministically as shall be demonstrated
for a simple case in the following.

After the introduction of a discrete dynamical time in classical physics
(3.1) as well as time resolution considerations of setting up and
measuring an experiment (4.1), the super position of two quantum states
is replaced by a fast harmonic oscillation between the states followed
by measuring it classically under the assumption of a discrete time
(4.2). In 4.3 the introduced concepts of discrete time and the fast
oscillation between eigenstates are applied to the singlet state of
two particles with spin $\frac{1}{2}$ without the request of entanglement
resolving thereby the EPR paradox, followed in paragraph 5 by a discussion.

\newpage{}

\section{Theory}

\subsection{The discreetness of time in classical physics }

It is assumed that in contrast to quantum mechanics with a continuous
time $t$, in classical physics time evolves in very small discrete
time steps $\Delta t\,$ of constant nature (for example $\Delta t\,$
could be the Planck time $\Delta t_{p}=\sqrt{\frac{h\,G}{2\pi\,c^{5}}}=\,5.4\,10^{-44}$
s with $h$ the Planck constant, $c$ the light velocity and $G$
the gravitational constant). This assumption is nourished on the finding
that the arrow of time and entropy can be derived by the introduction
of a discrete time {[}9{]} and on the request of a dual relationship
between energy and time from the corresponding uncertainty principle
$\Delta E\,\Delta t\,>h/2$ (i.e. since in quantum mechanics energy
is quantized time is (allowed to be) continuous and vice versa, since
in classical physics energy is continuous time is requested to be
discrete). Thus, by introducing a discrete time, any time-dependent
observable $A(t)$ is represented by a sequence of discrete values
(10,11):

\begin{equation}
(A_{0},\,t_{0}=0),\,(A_{1},\,t_{1}=1\,\Delta t),........,\,(A_{n},\,t_{n}=n\,\Delta t),......,\,(A_{N+1},\,t_{N+1}=[N+1]\,\Delta t)
\end{equation}
 with $(A_{0},\,t_{0})$ the initial and $(A_{N+1},\,t_{N+1})$ the
final measurement of the system under study and with with $n$ being
an element of the natural numbers including 0. Furthermore, the following
relationship between the continuous time in quantum mechanics $t$
and the discrete time in classical physics holds:

\begin{equation}
t_{n}=n\,\Delta t
\end{equation}

denoting $t_{n}$ to be the quantum mechanical time at classical time
point $n$. It is further assumed, that time is a single variable
(unlike space coordinates) and started at the beginning of the universe
and thus any object whether it is the system under study or the detector
used (i.e. the instrument used to measure the system) are with the
discrete time steps in tune/coherent to each other. 

In addition, it is important to mention that with state of the art
technologies the time precision of experiments and thus the time point
of measurement as well as the start of the experiment are many orders
of magnitude less accurate than $\Delta t$. Actually, since currently
the time resolution is in the order of $10^{-18}s$ (or less) and
assuming $\Delta t=\Delta t_{p}$, the starting time of the evolution
of the system as well as its end are not more precise than ca. $10^{26}\,\Delta t$.
Thus, each experiment is somewhat different and it is expected that
only an ensemble averaging over many experiments will give a valuable
result. With other words already because of these experimental limitations
the time-dependent deterministic quantum mechanics described by the
Schrödinger equation (3) can only in average calculate the outcome
of an experiment. This argumentation suggests that quantum mechanics
is well calculating a single outcome, but it is meaningless since
the experimental set up is not of sufficient quality (i.e. of sufficient
time resolution) and only upon ensemble averaging quantum mechanics
and experiments fit to each other.

\subsection{Modifying the super position of quantum states by a time-resolved
fast oscillation between the states}

Let us study a particle with spin $\frac{1}{2}$. In the standard
description of quantum mechanics (3), the eigenstates of a spin $\frac{1}{2}$
along the z-axis are the spin up $|\uparrow>_{z}$ state and the corresponding
orthogonal spin down state $|\downarrow>_{z}$ . When the particle
state is not defined (i.e. is not in one of the eigen states), it
is said that the particle is in both states, it is super imposed.
This can be described by the following wave function using the Dirac
notation:

\begin{equation}
\text{|\textgreek{Y}(t)>=\ensuremath{\frac{1}{\sqrt{2}}}\,}|\downarrow>_{z}+\frac{1}{\sqrt{2}}|\uparrow>_{z}
\end{equation}

While superimposed upon a single measurement, it is either in one
of the two states, and upon ensemble averaging of many measurements
it is 50\% in the spin up state and 50\% in the spin down state, expressed
as follows:

\begin{equation}
\text{<\textgreek{Y}(t)|\textgreek{Y}(t)>}_{z}=\frac{1}{2}<\downarrow|\downarrow>_{z}+\frac{1}{2}<\uparrow|\uparrow>_{z}
\end{equation}

. 

It is now proposed here to modify the super position concept by suggesting
that the spin $\frac{1}{2}$ system oscillates between the two states
harmonically (Figure 1) with either 

\begin{equation}
\text{|\ensuremath{\Psi_{hd}^{\downarrow}}(t)>}=\frac{1}{\sqrt{2}}\,cos(\frac{\pi\,t}{2\Delta t})|\downarrow>_{z}+\frac{1}{\sqrt{2}}\,sin(\frac{\pi\,t}{2\Delta t})|\uparrow>_{z}
\end{equation}
if at starting time the quantum state was in a spin down $|\downarrow>_{z}$
state,

or, with

\begin{equation}
\text{|\ensuremath{\Psi_{hd}^{\uparrow}}(t)>}=\frac{1}{\sqrt{2}}\,sin(\frac{\pi\,t}{2\Delta t})|\downarrow>_{z}+\frac{1}{\sqrt{2}}\,cos(\frac{\pi\,t}{2\Delta t})|\uparrow>_{z}
\end{equation}

if at the start of the experiment the quantum system was in a spin
up state $|\uparrow>_{z}$ . It is evident that with these wave function
descriptions denoted $\text{\textgreek{Y}}_{hd}(t)$ with $hd$ for
harmonic and discrete, there is no superposition because the system
oscillates forth and back between the two states with a periodicity
of $4\,\Delta t$.

As discussed above in measuring a single experiment one needs to consider
the imprecision in the starting and measurement time. Thus, it is
for a single measurement unknown whether the spin $\frac{1}{2}$ system
starts with spin up $|\uparrow>_{z}$ or spin down $|\downarrow>_{z}$.
Correspondingly, because with the current experimental technologies
the current time resolution of the measurement is many orders of magnitude
larger than the steps of the discrete time the measurement is done
at $t_{n}=n\,\Delta t$ with $n$being an even or odd number. An averaging
over many experiments needs to take into account these limitations.

Under these considerations let us assume the experiment did start
with the spin $\frac{1}{2}$ in the spin down state $|\downarrow>_{z}$
measuring

\begin{equation}
\text{<\ensuremath{\Psi_{hd}^{\downarrow}}(t)|\ensuremath{\Psi_{hd}^{\downarrow}}(t)>}_{z,\,n}=\frac{1}{2}\,cos^{2}(\frac{\pi\,n\Delta t}{2\Delta t})<\downarrow|\downarrow>_{z}+\frac{1}{2}\,sin^{2}(\frac{\pi\,n\Delta t}{2\Delta t})<\uparrow|\uparrow>_{z}
\end{equation}

(note the ``bra'' part may be regarded as the expectation value
of the measurement device) which results by measuring at a time $t_{n}$
with an even $n=2\,m$ (with $m$ being an element of the natural
numbers including 0) to

\begin{equation}
\text{<\ensuremath{\Psi_{hd}^{\downarrow}}(t)|\ensuremath{\Psi_{hd}^{\downarrow}}(t)>}_{z,\,2m}=\frac{1}{2}\,cos^{2}(\frac{\pi\,2m\Delta t}{2\Delta t})<\downarrow|\downarrow>_{z}=\frac{1}{2}<\downarrow|\downarrow>_{z}
\end{equation}

and for a time with an odd $n=2\,m+1$ to

\begin{equation}
<\Psi_{hd}^{\downarrow}(t)|\Psi_{hd}^{\downarrow}(t)>_{z,\,2m+1}=\frac{1}{2}\,sin^{2}(\frac{\pi\,(2m+1)\Delta t}{2\Delta t})<\uparrow|\uparrow>_{z}=\frac{1}{2}<\uparrow|\uparrow>_{z}
\end{equation}

Thus, the spin $\frac{1}{2}$ system that has started with a spin
down state $|\downarrow>$ is always detected as a spin up or a spin
down state although by our modified quantum mechanical description
there is an oscillation between the two states.

When ensemble averaged over many measurements with either a measurement
at even or odd time steps

\begin{equation}
<\text{\textgreek{Y}}_{hd}^{\downarrow}(t)|\text{\textgreek{Y}}_{hd}^{\downarrow}(t)>_{z}=\frac{1}{2}\text{<\ensuremath{\Psi_{hd}^{\downarrow}}(t)|\ensuremath{\Psi_{hd}^{\downarrow}}(t)>}_{z,\,2m}+\frac{1}{2}\text{<\ensuremath{\Psi_{hd}^{\downarrow}}(t)|\ensuremath{\Psi_{hd}^{\downarrow}}(t)>}_{z,\,2m+1}=\frac{1}{2}<\downarrow|\downarrow>_{z}+\frac{1}{2}<\uparrow|\uparrow>_{z}
\end{equation}

and thus 50\% of the case the system is in spin up and in 50\% of
the measurements it is in the spin down state, respectively.

Correspondingly, if the spin $\frac{1}{2}$ starts with the spin up
state $|\uparrow>$ it is detected as

\begin{equation}
\text{<\ensuremath{\Psi_{hd}^{\uparrow}}(t)|\ensuremath{\Psi_{hd}^{\uparrow}}(t)>}_{z,\,n}=\frac{1}{2}\,sin{}^{2}(\frac{\pi\,n\Delta t}{2\Delta t})<\downarrow|\downarrow>_{z}+\frac{1}{2}\,cos{}^{2}(\frac{\pi\,n\Delta t}{2\Delta t})<\uparrow|\uparrow>_{z}
\end{equation}

which results by measuring at a time $t_{n}$ with an even $n=2\,m$
(with $m$ being an element of the natural numbers including 0) to

\begin{equation}
\text{<\ensuremath{\Psi_{hd}^{\uparrow}}(t)|\ensuremath{\Psi_{hd}^{\uparrow}}(t)>}_{z,\,2m}=\frac{1}{2}\,cos^{2}(\frac{\pi\,2m\Delta t}{2\Delta t})<\uparrow|\uparrow>_{z}=\frac{1}{2}<\uparrow|\uparrow>_{z}
\end{equation}

and for a time with an odd $n=2\,m+1$ to

\begin{equation}
<\text{\textgreek{Y}}_{hd}^{\uparrow}(t)|\text{\textgreek{Y}}_{hd}^{\uparrow}(t)>_{z,\,2m+1}=\frac{1}{2}\,sin^{2}(\frac{\pi\,(2m+1)\Delta t}{2\Delta t})<\downarrow|\downarrow>_{z}=\frac{1}{2}<\downarrow|\downarrow>_{z}
\end{equation}

and under the condition that the spin system started with a spin up
state averaging over many measurements with either measurements at
even or odd time steps results in

\begin{equation}
<\text{\textgreek{Y}}_{hd}^{\uparrow}(t)|\text{\textgreek{Y}}_{hd}^{\uparrow}(t)>_{z}=\frac{1}{2}\text{<\ensuremath{\Psi_{hd}^{\uparrow}}(t)|\ensuremath{\Psi_{hd}^{\uparrow}}(t)>}_{z,\,2m}+\frac{1}{2}\text{<\ensuremath{\Psi_{hd}^{\uparrow}}(t)|\ensuremath{\Psi_{hd}^{\uparrow}}(t)>}_{z,\,2m+1}=\frac{1}{2}<\uparrow|\uparrow>_{z}+\frac{1}{2}<\downarrow|\downarrow>_{z}
\end{equation}

However, an ensemble measurement must take into account both the time
variation-induced variation in the starting state (i.e. either spin
up or spin down) as well as the measurement at even or odd time steps.
By doing so the following expression is obtained

\begin{equation}
\text{<\ensuremath{\Psi_{hd}}(t)|\ensuremath{\Psi_{hd}}(t)>}_{z}=\frac{1}{2}<\text{\textgreek{Y}}_{hd}^{\uparrow}(t)|\text{\textgreek{Y}}_{hd}^{\uparrow}(t)>_{z}+\frac{1}{2}<\text{\textgreek{Y}}_{hd}^{\downarrow}(t)|\text{\textgreek{Y}}_{hd}^{\downarrow}(t)>_{z}
\end{equation}
\[
=\frac{1}{2}<\downarrow|\downarrow>_{z}+\frac{1}{2}<\uparrow|\uparrow>_{z}=<\text{\textgreek{Y}}(t)|\text{\textgreek{Y}}(t)>_{z}
\]

As demonstrated in eq. 15. the same result as in the conventional
quantum mechanics approach is obtained but without requesting a simultaneous
superposition of the two states generating a modified quantum mechanics
that is upon exact measurement deterministic, while under current
experimental time accuracies predicts statistically the outcome of
an experiment identical to standard quantum mechanics. 

\includegraphics[scale=0.3]{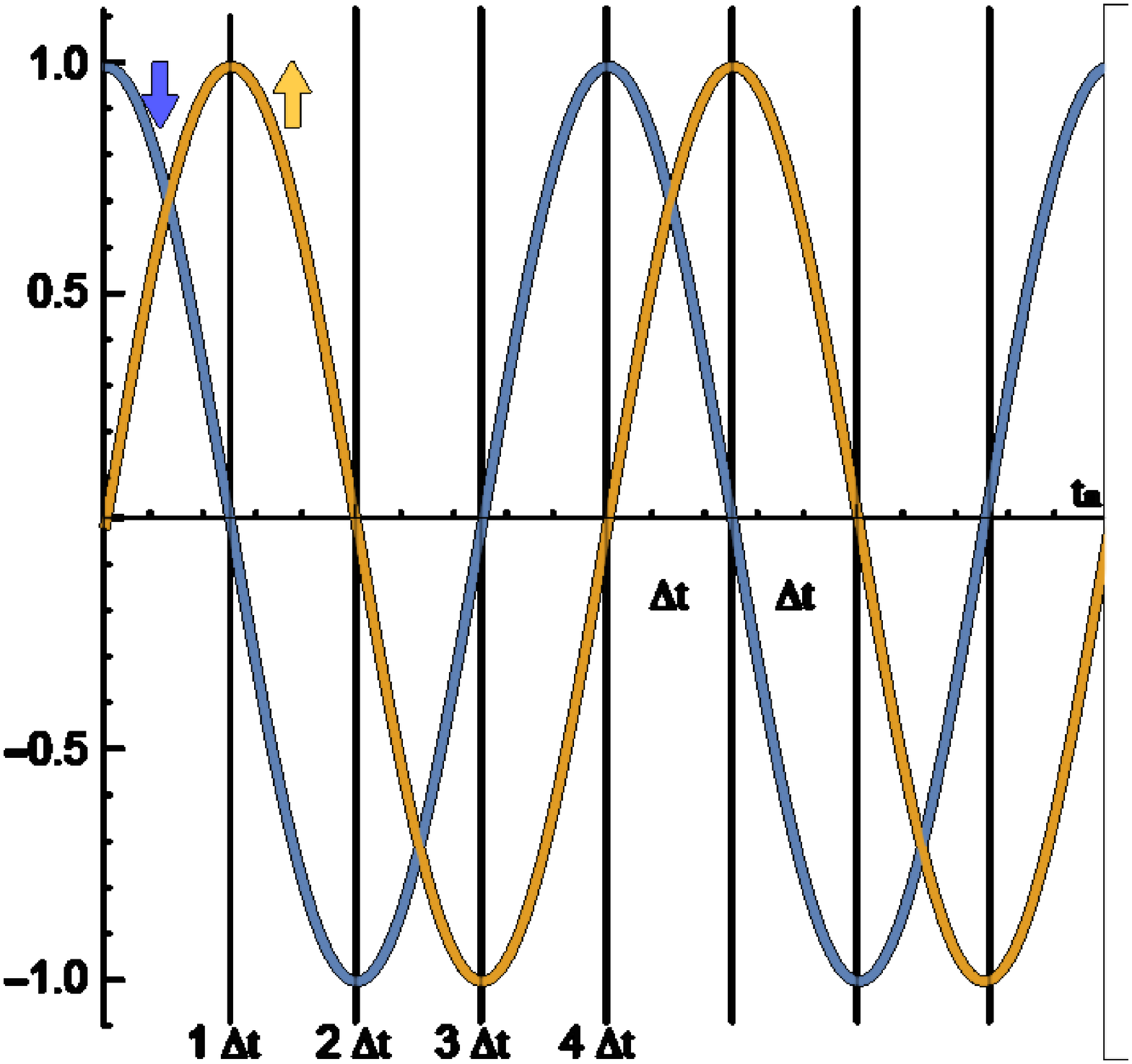}

Figure 1: The time-resolved harmonic oscillation between spin down
$|\downarrow>_{z}$ and spin up $|\uparrow>_{z}$ of a spin $\frac{1}{2}$
system is illustrated. With grey lines are the possible time points
of the measurement indicated. They are in time steps of $\Delta t$
as labeled.

\subsection{The Einstein-Podolsky-Rosen paradox of a singlet state studied under
discrete time physics }

An entangled quantum system is discussed composed of two spins $\frac{1}{2}$
particles in a singlet state. This means that since the sum of the
spin is 0, either of the particles is in the spin up state and the
other in the spin down state, or vice versa, respectively. This singlet
state can be described using the standard quantum mechanics formulation
as follows:

\begin{equation}
\text{|\textgreek{Y}(t)>}=\frac{1}{\sqrt{2}}|\uparrow\downarrow>_{z}-\frac{1}{\sqrt{2}}|\downarrow\uparrow>_{z}
\end{equation}

Please note, that because of the entanglement the wave function of
the singlet state is not the product of the two individual ones $\text{|\ensuremath{\Psi_{i}}(t)>}=\frac{1}{\sqrt{2}}|\downarrow_{i}>_{z}+\frac{1}{\sqrt{2}}|\uparrow_{i}>_{z}\:with\:i=1,2$. 

If the entangled state is made observable along the z-axis

\begin{equation}
<\text{\textgreek{Y}}(t)|\text{\textgreek{Y}}(t)>_{z\,}=\frac{1}{2}<\uparrow\downarrow|\uparrow\downarrow>_{z}+\frac{1}{2}<\downarrow\uparrow|\downarrow\uparrow>_{z}=\frac{1}{2}<\uparrow_{1}|\uparrow_{1}>_{z}<\downarrow_{2}|\downarrow_{2}>_{z}+\frac{1}{2}<\downarrow_{1}|\downarrow_{1}>_{z}<\uparrow_{2}|\uparrow_{2}>_{z}
\end{equation}

which means that if particle 1 is upon measurement in the spin up
state, particle 2 is simultaneously in the spin down state and vice
versa. After many measurements, it is further the finding that in
50\% of the cases particle 1 is in the spin up state and particle
2 is in the spin down state, while in the other 50\% of measurements
particle 1 is in the spin down state and particle 2 is in the spin
up state, respectively. 

For a more comprehensive analysis of the system under study, particle
1 is first detected along the z-axis followed by the measurement of
particle 2 along an arbitrary angle $\varphi$to the z-axis {[}3,
4{]}. If the first measurement results in $<\uparrow_{1}|\uparrow_{1}>_{z}$
the spin component of the wave function of particle 2 is given by
$|\downarrow>_{z}=sin\,\,\frac{\varphi}{2}|\uparrow_{2}>_{\varphi}+cos\,\,\frac{\varphi}{2}|\downarrow_{2}>_{\varphi}$
. The probability that the result of the second experiment is also
positive (i.e. $<\uparrow_{2}|\uparrow_{2}>_{\varphi}$ ) is therefore
given by $P_{++}(\varphi)=sin^{2}\frac{\varphi}{2}$ and correspondingly
the probability that spin 2 is measured as a spin down state $<\downarrow_{2}|\downarrow_{2}>_{\varphi}$
is $P_{+-}(\varphi)=cos^{2}\frac{\varphi}{2}$. (3). Accordingly,
if the first measurement is $<\downarrow_{1}|\downarrow_{1}>_{z}$
the probability that spin two is also negative is given by $P_{--}(\varphi)=sin^{2}\frac{\varphi}{2}$
and correspondingly the probability that spin 2 is measured as a spin
up state is $P_{-+}(\varphi)=cos^{2}\frac{\varphi}{2}$, respectively
(3). 

In contrast to standard quantum mechanics described above, by using
the harmonic oscillation and discrete time concept introduced above,
the singlet state can be described by the product of the individual
states $|\text{\textgreek{Y}}_{i,hd}(t)>$ as follows: 

\begin{equation}
|\;\text{\textgreek{Y}}_{hd}(t)^{\downarrow\uparrow}>=|\text{\textgreek{Y}}_{1,hd}^{\downarrow}(t)>*|\text{\textgreek{Y}}_{2,hd}^{\uparrow}(t)>
\end{equation}

if particle 1 started with spin down and with 

\begin{equation}
|\,\text{\textgreek{Y}}_{hd}(t)^{\uparrow\downarrow}>=|\text{\textgreek{Y}}_{1,hd}^{\uparrow}(t)>*|\text{\textgreek{Y}}_{2,hd}^{\downarrow}(t)>
\end{equation}

if particle 1 started with spin up, respectively. Using the formulations
of eqs. 5 and 6 these equations can be rewritten to 

\begin{equation}
|\text{\textgreek{Y}}_{hd}(t)^{\downarrow\uparrow}>=\frac{1}{2}\{[cos(\frac{\pi\,t}{2\Delta t})|\downarrow_{1}>_{z}+\,sin(\frac{\pi\,t}{2\Delta t})|\uparrow_{1}>_{z}]*[sin(\frac{\pi\,t}{2\Delta t})|\downarrow_{2}>_{z}+\,cos(\frac{\pi\,t}{2\Delta t})|\uparrow_{2}>_{z}]
\end{equation}

\[
=\frac{1}{2}cos^{2}(\frac{\pi\,t}{2\Delta t})|\downarrow_{1}>_{z}|\uparrow_{2}>_{z}+\frac{1}{2}sin^{2}(\frac{\pi\,t}{2\Delta t})|\uparrow_{1}>_{z}|\downarrow_{2}>_{z}+\frac{1}{2}cos(\frac{\pi\,t}{2\Delta t})\,sin(\frac{\pi\,t}{2\Delta t})|\downarrow_{1}>_{z}|\downarrow_{2}>_{z}
\]
\[
+\frac{1}{2}sin(\frac{\pi\,t}{2\Delta t})\,cos(\frac{\pi\,t}{2\Delta t})|\uparrow_{1}>_{z}|\uparrow_{2}>_{z}
\]

and

\[
\text{\textgreek{Y}}_{hd}(t)^{\uparrow\downarrow}>=\frac{1}{2}[cos(\frac{\pi\,t}{2\Delta t})|\uparrow_{1}>_{z}+\,sin(\frac{\pi\,t}{2\Delta t})|\downarrow_{1}>_{z}]*[sin(\frac{\pi\,t}{2\Delta t})|\uparrow_{2}>_{z}+\,cos(\frac{\pi\,t}{2\Delta t})|\downarrow_{2}>_{z}]\}
\]

\begin{equation}
\text{|\ensuremath{\text{\textgreek{Y}}_{hd}}(t\ensuremath{)^{\uparrow\downarrow}}>}=\frac{1}{2}[cos^{2}(\frac{\pi\,t}{2\Delta t})|\uparrow_{1}>_{z}|\downarrow_{2}>_{z}+\frac{1}{2}[sin^{2}(\frac{\pi\,t}{2\Delta t})|\downarrow_{1}>_{z}|\uparrow_{2}>_{z}+\frac{1}{2}cos(\frac{\pi\,t}{2\Delta t})\,sin(\frac{\pi\,t}{2\Delta t})|\uparrow_{1}>_{z}|\uparrow_{2}>_{z}
\end{equation}

\[
+\frac{1}{2}sin(\frac{\pi\,t}{2\Delta t})cos(\frac{\pi\,t}{2\Delta t})|\downarrow_{1}>_{z}|\downarrow_{2}>_{z}
\]

Please note, that the last terms with the mixed $sin$ and $cos$
functions are never observable (for any $t_{n}=n\,\Delta t\,$) because
$sin(\frac{\pi\,t}{2\Delta t})cos(\frac{\pi\,t}{2\Delta t})=\frac{1}{2}sin(\frac{\pi\,t}{\Delta t})$.
These terms although present are not detectable because they oscillate
twice as fast and thus at any measurement time are 0. It is important
to notice that at the origin of the doubling of the oscillation is
the request for a single time variable: both particles are evolving
with the same time and are thus connected to each other, while not
entangled. Thus, if a meaurement is made on particle 1 at a given
time point it has also consequences for particle 2. For example, if
particle 1 started with spin down (i.e. eq. 20) and in the measurement
is observed to have spin down $<\downarrow_{1}|\downarrow_{1}>_{z}$
(i.e. $n=2m$ with $m$ being an element of the natural numbers including
0)

\begin{equation}
<\downarrow_{1}\text{|\ensuremath{\text{\textgreek{Y}}_{hd}}(t\ensuremath{)^{\downarrow\uparrow}}>}_{z,2m}=\frac{1}{2}<\downarrow_{1}|\downarrow_{1}>_{z}|\uparrow_{2}>_{z}
\end{equation}
for particle 2 the spin up state $<\uparrow_{2}|\uparrow_{2}>_{z}$
will and must be detected and any time point later. 

The same result is obtained if particle 1 started with spin up (eq.
21) and the measurement is performed at a time point $t_{n}$ with
$n=2m+1$

\begin{equation}
<\downarrow_{1}\text{|\ensuremath{\text{\textgreek{Y}}_{hd}}(t\ensuremath{)^{\uparrow\downarrow}}>}_{z,2m+1}=\frac{1}{2}<\downarrow_{1}|\downarrow_{1}>_{z}|\uparrow_{2}>_{z}
\end{equation}

Alternatively, if particle 1 is started with spin down (i.e. eq. 20)
and the measurement of particle 1 is performed at a time point $t_{n}$
with $n=2m+1$ the spin up state $<\uparrow_{1}|\uparrow_{1}>_{z}$
is observed

\begin{equation}
<\uparrow_{1}\text{|\ensuremath{\text{\textgreek{Y}}_{hd}}(t\ensuremath{)^{\downarrow\uparrow}}>}_{z,2m+1}=\frac{1}{2}<\uparrow_{1}|\uparrow_{1}>_{z}|\downarrow_{2}>_{z}
\end{equation}

and thus for particle 2 the spin down state $<\downarrow_{2}|\downarrow_{2}>_{z}$
will be detected at any time point later. 

The same result is obtained if particle 1 is started with spin up
(i.e. eq. 20) and the measurement of particle 1 is performed at a
time point $t_{n}$ with $n=2m$ yielding the spin up state $<\uparrow_{1}|\uparrow_{1}>_{z}$
: $<\uparrow_{1}\text{|\ensuremath{\text{\textgreek{Y}}_{hd}}(t\ensuremath{)^{\uparrow\downarrow}}>}_{z,2m}=\frac{1}{2}<\uparrow_{1}|\uparrow_{1}>_{z}|\downarrow_{2}>_{z}$

In analogy (i.e by exchanging the numbering of the spins in the formulas
above) the corresponding results are obtained if first spin 2 is detected.

Finally, if averaged over many measurements in 50\% of the measurements
particle 1 is found in the spin up state and particle 2 in the down
state, and vice versa in the other 50\% of the measurements particle
1 is found in the spin down state and particle 2 in the spin up state,
respectively:

\begin{equation}
\text{<\text{\ensuremath{\Psi_{hd}}(t)}|\ensuremath{\Psi_{hd}}(t)>}_{z}=\frac{1}{2}<\uparrow_{1}|\uparrow_{1}>_{z}<\downarrow_{2}|\downarrow_{2}>_{z}+\frac{1}{2}<\downarrow_{1}|\downarrow_{1}>_{z}<\uparrow_{2}|\uparrow_{2}>_{z}=<\text{\ensuremath{\Psi}(t)}|\ensuremath{\Psi}(t)>_{z}
\end{equation}

Thus, under the assumption of a single coherent time variable, which
is discrete in classical physics and by replacing the quantum super
position by a fast oscillation between the states the detection along
the z-axis of a singlet state can be described without the request
of quantum entanglement. Actually, the quantum entanglement as exemplified
here by the singlet state can be explained as two individual systems
that are coherent in time and thus non-locally connected since it
is assumed that there is only one single time variable. 

This finding holds also if after the measurement of particle 1 along
the z-axis at time point $t_{n}$, the second particle is measured
at an angle $\varphi$ to the z-axis at a later time point $t_{n+k}=t_{n}+k\,\Delta t$
(with $k$ being an element of the natural numbers including 0). It
is important to notice that the request for a single time variable
is thereby considered. Let us first study the case that particle 1
is measured at time point $t_{n}$ to have spin down $<\downarrow_{1}|\downarrow_{1}>_{z}$
(i.e. either eqs. 22 or 23). For preparing the measurement of particle
2 along the $\varphi$axis a time interval $\delta_{k}=k\,\Delta t$
later the following expression is obtained

\begin{equation}
<\downarrow_{1}\text{|\ensuremath{\Psi_{hd}}(t)>}_{z}=\frac{1}{2}<\downarrow_{1}|\downarrow_{1}>_{z}|\uparrow_{2}>_{z}=\frac{1}{2}<\downarrow_{1}|\downarrow_{1}>_{z}[cos(\frac{\pi\,\delta_{k}}{2\Delta t})\,cos\,\,\frac{\varphi}{2}|\uparrow_{2}>_{\varphi}+sin(\frac{\pi\,\delta_{k}}{2\Delta t})\,sin\,\frac{\varphi}{2}|\downarrow_{2}>_{\varphi}]
\end{equation}

The measurement of particle 2 at a time interval $\delta_{k}=k\,\Delta t$
later with an even $k=2\,m'$ (with $m'$ being an element of the
natural numbers including 0).

\begin{equation}
\text{<\text{\ensuremath{\Psi_{hd}}(t)}|\ensuremath{\Psi_{hd}}(t)>}_{z,\varphi,2m'}=\frac{1}{2}<\downarrow_{1}|\downarrow_{1}>_{z}\,cos^{2}\,\,\frac{\varphi}{2}<\uparrow_{2}|\uparrow_{2}>_{\varphi}
\end{equation}

and correspondingly the measurement at $\delta_{k}=k\,\Delta t$ later
with an odd $k=2\,m'+1$ yields

\begin{equation}
\text{<\text{\ensuremath{\Psi_{hd}}(t)}|\ensuremath{\Psi_{hd}}(t)>}_{z,\varphi,2m'+1}=\frac{1}{2}<\downarrow_{1}|\downarrow_{1}>_{z}\,sin^{2}\,\,\frac{\varphi}{2}<\downarrow_{2}|\downarrow_{2}>_{\varphi}
\end{equation}

The corresponding results are obtained if the particle 1 is measured
at time point $t_{n}$ to have spin up $<\uparrow_{1}|\uparrow_{1}>_{z}$
(i.e. eq. 24 and thereafter) and the measurement of particle 2 along
the $\varphi$axis a time interval $\delta_{k}=k\,\Delta t$ later
is prepared

\begin{equation}
<\uparrow_{1}\text{|\ensuremath{\Psi_{hd}}(t)>}_{z}=\frac{1}{2}<\uparrow_{1}|\uparrow_{1}>_{z}|\downarrow_{2}>_{z}=\frac{1}{2}<\uparrow_{1}|\uparrow_{1}>_{z}[sin(\frac{\pi\,\delta_{k}}{2\Delta t})\,sin\,\,\frac{\varphi}{2}|\uparrow_{2}>_{\varphi}+cos(\frac{\pi\,\delta_{k}}{2\Delta t})\,cos\,\frac{\varphi}{2}|\downarrow_{2}>_{\varphi}]
\end{equation}
The measurement of particle 2 at $\delta_{k}=k\,\Delta t$ later with
an $k=2\,m'$ yields
\begin{equation}
\text{<\text{\ensuremath{\Psi_{hd}}(t)}|\ensuremath{\Psi_{hd}}(t)>}_{z,\varphi,2m'}=\frac{1}{2}<\uparrow_{1}|\uparrow_{1}>_{z}\,cos^{2}\,\,\frac{\varphi}{2}<\downarrow_{2}|\downarrow_{2}>_{\varphi}
\end{equation}

and correspondingly if measured at $\delta_{k}=k\,\Delta t$ later
with $k=2\,m'+1$ yields

\begin{equation}
\text{<\text{\ensuremath{\Psi_{hd}}(t)}|\ensuremath{\Psi_{hd}}(t)>}_{z,\varphi,2m'+1}=\frac{1}{2}<\uparrow_{1}|\uparrow_{1}>_{z}\,sin^{2}\,\,\frac{\varphi}{2}<\uparrow_{2}|\uparrow_{2}>_{\varphi}
\end{equation}

When these individual measurements are ensemble averaged over many
experiments the same probabilities are obtained as calculated by standard
quantum mechanics:

\begin{equation}
\text{<\text{\ensuremath{\Psi_{hd}}(t)}|\ensuremath{\Psi_{hd}}(t)>}_{z,\varphi}=\frac{1}{2}cos^{2}\,\,\frac{\varphi}{2}[<\uparrow_{1}|\uparrow_{1}>_{z}<\downarrow_{2}|\downarrow_{2}>_{\varphi}+<\downarrow_{1}|\downarrow_{1}>_{z}<\uparrow_{2}|\uparrow_{2}>_{\varphi}]
\end{equation}

\[
+\frac{1}{2}sin^{2}\,\,\frac{\varphi}{2}[<\uparrow_{1}|\uparrow_{1}>_{z}\,<\uparrow_{2}|\uparrow_{2}>_{\varphi}+<\downarrow_{1}|\downarrow_{1}>_{z}\,<\downarrow_{2}|\downarrow_{2}>_{\varphi}]=<\text{\ensuremath{\Psi}(t)}|\ensuremath{\Psi}(t)>_{z,\varphi}
\]

\section{Discussion}

In addition to a quantized energy, quantum mechanics differs from
classical physics by several counter intuitive properties such as
the super position of quantum states, the non local nature of quantum
mechanics, and the lack of the possibility to calculate a single measurement.
The presented approach of combining the presumed existence of a discrete
time in classical physics coherently connected to a single continuous
time variable in quantum mechanics and the fast oscillation of the
quantum system between its eigenstates is able to resolve some of
these phenomena. It localizes the problem of measuring a quantum state
(i.e. collapse of the wave function) to the issue of exact timing
and explains why quantum mechanics is not able to calculate single
experimental outcomes at current time resolutions, but calculates
them correctly in average. It further resolves the causality issue
as outlined in the EPR paradox by Einstein and coworkers (1). In this
context it is noted that a Gedankenexperiment by A. Suarez further
shows that while standard quantum mechanics is non local, it is only
so by requesting relativity, which is local in nature, a dilemma which
can be resolved by the introduction of a discrete time (11). At the
root of the causal nature of the presented modified quantum mechanics
theory is thereby in addition the request of a single time variable,
which is believed to be a logic axiom for a theory that guarantees
causality. 

While the presented modification of quantum mechanics for a two spin
system may be regarded sound, the generalization - how it can be translated
into a system with many eigen states - remains to be established.
One may speculate that the system goes periodically through all the
eigen states (please note, the mathematical description of such a
system may include step functions, which may result in other demanding
mathematical reformulations such as the Schrödinger equation). 

Furthermore, finding experimental support for the presence of a discrete
time and the modified quantum mechanics appear to be difficult since
the current experimental time resolution is ca 26 orders of magnitude
away from the Planck time $\Delta t_{p}$. Nonetheless, it can be
stated, that the better the time resolution of the experiment (and
eventually also the faster the processes under study) the more deviation
from quantum mechanics towards a classical/deterministic behavior
of the system is expected. Alternatively, the nature of the break
down of a quantum mechanical system into its classical analog may
give valuable hints in favor or against the presented theory. For
example if a quantum system oscillates through all the eigen states
in steps of $4\,\Delta t_{p}$ as speculated above, the system may
behave classically or/and deterministic if the time the system needs
to go through all significant eigen states is longer than the experimental
time resolution (i.e. if $f*4\,\Delta t_{p}$ is larger than the time
resolution of the experiment with $f$ being the number of significant
eigen states of the system), while it would behave quantum mechanically
if the time resolution of the experiment would be less (please note
at the current time resolution this requests the study of a system
with ca $10^{26}$ significantly populated eigen states). With other
words, if time is discrete the presented theory is able to calculate
the size dependence of a system to behave as either a quantum or classical
system. 

It is evident that the concept of a discrete time is at the root of
the presented interpretation. It is however not a very popular concept,
although upon energy quantization in quantum mechanics the introduction
of a discrete time appears to be obvious (12-17). Of course, in the
presented manuscript, only the simplest examples have been looked
at yet and further considerations are necessary to show generality
(see also discussion above). However the recent success in the derivation
of a microscopic entropy under the assumption of a discrete time strengthens
the hypothesis that time in classical physics may be discrete (10)
- a possibility that is worth to be disputed further.

\section{Acknowledgment}

We would like to thank the ETH for unrestrained financial support
and Dr. Gianni Blatter, Dr. Alexander Sobol, and Dr. Witek Kwiatkowski
for helpful discussions.

\section{References}

1. Einstein A, Podolsky B, Rosen N 1935 Can Quantum-Mechanical Description
of Physical Reality Be Considered Complete? \textit{Phys. Rev. }47:777\textendash 780. 

2. Schrödinger E 1935 Die gegenwärtige Situation in der Quantenmechanik.
\textit{Naturwissenschaften} 23:823807\textendash 828812.

3. Greiner W, Neise L, Stöcke H 1989 \textit{Quantenmechanik}. Verlag
Harri Deutsch: Frankfurt am Main.

4. Bell J 1964 On the Einstein Podolsky Rosen Paradox. \textit{Physics}
1:195\textendash 200.

5. Aspect A(1999 Bell's inequality test: more ideal than ever, \textit{Nature}
398:189\textendash 90.

6. Hense B, Bernien H, Dreau AE, Reiserer, A, Kalb N, Blok MS, Ruitenberg,
J, Vermeulen RFL, Schouten RN, Abellan C, Amya W, Prueri V, Mitchell
MW, Markham M, Twitchen DJ, Ekouss D, Wehner S, Taminiau TH, Hanson
R 2015 Loophole-free Bell inequality violation using electron spins
separated by 1.3 kilometres, \textit{Nature} 526:682-686.

7. Wittmann B, Ramelow S, Steinlechner F, Langford NK, Brunner N,
Wiseman HM, Ursin R, Zeilinger A 2012 Loophole-free Einstein\textendash Podolsky\textendash Rosen
experiment via quantum steering N. \textit{J. of Physics} 14, 053030-05042.

8. Bohm D 1952 A Suggested Interpretation of the Quantum Theory in
Terms of 'Hidden Variables' I. \textit{Physical Review} 85: 66\textendash 179.

9. Schrödinger E, Born M 1935 Discussion of probability relations
between separated systems. \textit{Mathematical Proceedings of the
Cambridge Philosophical Society} 31:555\textendash 563.

10. Riek R 2014 A derivation of a microscopicentropy and time irreversibility
from the discreteness of time. \textit{Entropy} 16:3149-3172.

11. Suarez A 2010 Unified demonstration of nonlocality at detection
and the Michelson-Morley result, by a single-photon experiment. \textit{arXiv:1008.3847}

12. Lee TD 1983 Can time be a discrete dynamical variable? \textit{Physics
Letters} 122B:217-220.

13. Levi R 1927 Théorie de l'action universelle et discontinue \textit{Journal
de Physique et le Radium} 8:182-198. 

14. Caldirola P 1980 The introduction of the chronon in the electron
theory and a charged lepton mass formula. \textit{Lett. Nuovo Cim}.
27: 225\textendash 228.

15. Jaroszkiewicz G and Norton K 1997 Principles of discrete time
mechanics: I. Particle systems \textit{J. Phys. A. Math. Gen} 30:3115-3144.

16. Jaroszkiewicz G and Norton K 1997 Principles of discrete time
mechanics: II. Classical field theory \textsl{J. Phys. A. Math. Gen}
30:3145-3164.

17. Farias RAH and Recami E 1997 Introduction f a quantum of time
(``chronon'') and its Consequences for Quantum Mechanics. a\textit{rXiv:quant-ph/9706059}
\end{document}